\documentclass[english,aps,preprint]{revtex4}
\usepackage[T1]{fontenc}
\usepackage[latin9]{inputenc}

\makeatletter

\providecommand{\LyX}{L\kern-.1667em\lower.25em\hbox{Y}\kern-.125emX\@}

\makeatother

\usepackage{babel}

\begin{document}

\preprint{This line only printed with preprint option}

\title{Self collimation of ultrasound in a three-dimensional sonic crystal}

\author{Ester Soliveres, Víctor Espinosa, Isabel Pérez-Arjona, Víctor J.
Sánchez-Morcillo}

\email{iparjona@upvnet.upv.es}

\homepage{http://www.upv.es/entidades/IGIC}

\affiliation{Institut d'Investigació per a la Gestió Integrada de Zones Costaneres,
Universitat Politècnica de València, Crta. Natzaret Oliva s/n, 46730
Grau de Gandia, Spain}

\author{Kestutis Staliunas}

\email{kestutis.staliunas@icrea.es}

\affiliation{Institució Catalana de Recerca i Estudis Avancats (ICREA), Departament
de Física i Enginyeria Nuclear, Universitat Politècnica de Catalunya,
Colom 11, 08222 Terrassa, Spain }
\begin{abstract}
We present the experimental demonstration of self-collimation (subdiffractive
propagation) of an ultrasonic beam inside a three-dimensional (3D)
sonic crystal. The crystal is formed by two crossed steel cylinders
structures in a woodpile-like geometry disposed in water. Measurements
of the 3D field distribution show that a narrow beam which diffractively
spreads in the absence of the sonic crystal is strongly collimated
in propagation inside the crystal, demonstrating the 3D self-collimation
effect. 
\end{abstract}
\maketitle

Wave beams diverge when they propagate in homogeneous materials due
to diffraction. Nevertheless, a particular regime where diffraction
spreading vanishes, the so-called self-collimation, was predicted
a decade ago for electromagnetic waves propagating in optically periodic
materials (photonic crystals) \cite{Kosaka99}. Inside a photonic
crystal the dispersion relations for propagation (Bloch) modes are
modified, and the envelopes of electromagnetic waves can propagate
without diffraction broadening \cite{witzens,chigrin}. The self-collimation
effect has been studied not only for electromagnetic, but also for
the other kind of waves. The effect analogous to the self-collimation
has been recently predicted for the matter waves\cite{Bose-einstein}.
Also subdiffractive propagation of sonic beams inside the phononic
(or sonic) crystals was predicted \cite{2DTeor}. The vanishing of
diffraction in the wave propagation along periodic crystals has been
so far experimentally demonstrated for electromagnetic waves in optical
\cite{Kosaka99,rakich} and microwave \cite{Lu06} frequencies, and
recently, for the ultrasonic beam propagation inside a sonic crystal
\cite{2DExp}. However, most of the beam propagation effects, in particular
the self-collimation effect on which we focus in this work, have been
addressed mainly in two-dimensional (2D) systems. The three-dimensional
(3D) systems are more complicated not only for the experimental study
but also in the numerical level, where the FDTD calculations are extremely
time consuming. From the experimental point of view, to the best of
our knowledge, the 3D self-collimation has been observed only for
microwaves \cite{Lu06} but never for the optical frequencies. Also
the 3D self-collimation has never been experimentally demonstrated
for the other than electromagnetic waves, i.e. the matter waves, or
the sound waves. Here we demonstrate the 3D self-collimation effect
in acoustics, i.e. the nondifractive propagation of an ultrasonic
beam through a 3D sonic crystal. The sonic crystal used in the experiment
can be considered as formed by two squared 2D structures like that
studied in \cite{2DExp}, rotated by 90 degrees and interlaced one
into another (Figure 1). Each of them are formed by 20x20 steel cylinders
with a radius r = 0.8 mm. The lattice constant is $a_{x}$ = $a_{y}$
= $a$ = 5.25 mm, where $a_{x}$, $a_{y}$ are the spatial periods
along x and y direction respectively. The beam is propagated along
the z direction inside the crystal. As the radius of the cylinders
r is smaller than the shift between the two interlaced structures,
$a$/2 in z direction, it results in a contact-free woodpile geometry.
This differs from the previously studied 3D sonic crystals, where
scatterers are located forming cubic lattices with face centered cubic
(fcc), body centered cubic (bcc) and simple cubic symmetries \cite{kuang06}.
Contrary to the most common configuration in experiments using liquid-solid
crystals, where in contact solid spheres are used as scatterers \cite{yang04},
the main advantage of the contact-free crystal is its relatively large
``transparency'' reducing the energy losses in propagation and in
the interfaces of the crystal. 

The experimental setup consists of a source of ultrasonic wave, the
above described 3D periodic structure and a needle hydrophone (to
measure the acoustic field); all these components are immersed in
a plexiglass tank filled with distilled water. The frequency tunable
source is a piezoelectric-based commercial projector with a resonant
frequency at 192 kHz, that can be tuned to a range of frequencies
belonging to the second propagation band in of our crystal (200 to
260 KHz) where the experiment is performed. The needle hydrophone
is an Onda Corp. HNR-0500 with a calibrated operating band between
0.25 and 10 MHz, that can be used for stable relative measurements
in the frequency range of interest. The acoustic signals are generated
and captured using a PXI National Instruments system with synchronized
signal generator and oscilloscope cards with oversampling frequency
capability. To position the hydrophone, three motorized axis are governed
by the acquisition system, mapping the acoustical beam. The excited
signal is a tone burst of several cycles of the studied frequency;
the pulse is long enough to assume the CW propagation inside the crystal
but sufficiently short to discard unwanted reflections coming from
the tank walls in the captured signal. The measured pressure levels
are low, assuring the linear regime. In order to reduce the noise
the temporal averaging of several pulses (up to 20) is performed and
a Butterworth eighth order band-pass filter is applied. The 3D sonic
crystal, as explained above, can be considered as two interlaced 2D
crystals studied in \cite{2DTeor,2DExp}. In the 2D case the nondiffractive
propagation was predicted for two different frequencies in the first
and second bands respectively, and experimentally verified for the
second band (225 kHz). Here, in the present letter, we study experimentally
the propagation through the 3D crystal in the second band by varying
the frequency around the above mentioned 2D self-collimation frequency
(225 kHz). 

Figure 2 shows the effect of the crystal on the propagation of the
beam. Fig. 2(a) shows the sound intensity distributions in the transversal
planes just after transducer, i.e. at a distance of 3 mm from the
transducer plane. Fig. 2(b) shows the distributions in a free (without
crystal) propagation over the distance of 115 mm, respectively. The
propagated beam is slightly anisotropic, i.e. is slightly broader
in the horizontal direction because the adapting layer of the emitting
transducer has a certain curvature (astigmatism) in that plane, acting
as a cylindrical diverging lens. Fig. 2(c) shows the amplitude profile
of the sound beam at the rear face of the crystal, measured at the
same 115 mm distance as in Fig. 2(b). The diameter of the central
part of the beam remains nearly of the same order than the input (just
slightly broadened), clearly indicating the effect of self-collimation.
Besides the central self-collimated beam, the side-lobes appear which
correspond to the diverging wave vectors. The side-lobes are related
with the excitation of the additional Bloch modes (in addition to
the basic subdifractive one), and require a separate study. We note
that these side-lobes disappear after the larger distances behind
the sonic crystal (not shown). Fig. 3 depicts the variation of the
beam width (measured at half amplitude) versus frequency. A minimum
is reached for the self-collimation frequency of the 2D case \cite{2DExp}.
The width is normalized here to the width of the beam after propagating
the same distance without crystal, and computed subtracting the noise
level amplitude with no interpolation between the spatially sampled
points. The depicted width is the average of the different measurements
of the width for the different vertical cuts of the transversal plane. 

The continuous curve in Fig. 3 is the theoretical fit. In paraxial
treatment, the beam broadening after the propagation in a homogeneous
material with diffraction coefficient $d$ is given by $\Delta x^{2}(z)=\Delta x_{0}^{2}+(4\, d\, z)^{2}/\Delta x_{0}^{2}$,
where $\Delta x_{0}$ is the initial width and $z$ the propagation
distance. This classical formula for the propagation of Gaussian beams
can be extended to the case of inhomogeneous media, taking into account
the particular dependence of the diffraction coefficient with the
frequency. In our case, we can assume to a first order approximation
that the diffraction coefficient depends linearly with frequency,
and that changes the sign (it cancels) at the self-collimation point.
The other coefficients are selected as those with a better fit to
the experimental data. We note that our 3D crystal is in fact the
enlace of two orthogonally oriented 2D structures, therefore the spatial
modulation $m$ of the acoustic parameters (bulk modulus and density)
is the additive function of corresponding 2D crystals: $m\left(x,y,z\right)=m_{1}\left(x,z\right)+m_{2}\left(y,z\right)$
, where the both functions depend separately on x and y. Under paraxial
approximation, where the field propagation is described by the Schrödinger
equation, the acoustic field (pressure field, $p$) factorize in this
case of additive potentials: $p\left(x,y,z\right)=p_{1}\left(x,z\right)\times p_{2}\left(y,z\right)$.
The 3D propagation problem then simplifies into two independent 2D
propagation problems studied before in \cite{2DTeor} and \cite{2DExp}.
Strictly speaking the propagation in the studied by us 3D sonic crystal
is not paraxial, since the sound waves diffract at relatively large
angles in every slice. However, having in mind that the self-collimation
phenomenon is usually analogous in paraxial and nonparaxial cases
(see e.g.\cite{Loiko&kestas} for comparison of the self-collimation
in these cases) the similarity of the measured sound profiles in 3D
case with product of 2D profiles \cite{2DExp}, as well as the good
coincidence between the self-collimation frequencies in 3D and 2D
cases is plausible. In conclusion, a non-contact woodpile sonic crystal,
consisting of two interlaced 2D structures of periodic arrays of steel
cylinders in water, has been fabricated. Experiments of ultrasonic
beam propagation through such crystal have demonstrated the self-collimation
propagation in a 3D sonic crystal. Two additional remarks should be
done about the advantages of the used crystal: on one hand, the crystal
differs from the most common close-packed solid spheres crystals in
the sense that, in the present crystal the filling fraction can be
arbitrarily modified allowing to design the crystal without the restriction
existent in close-packed spheres, where the filling fraction is fixed
at $f_{max}=0.7405$. In our case, the filling factor is $f=147$,
i.e. twice than in the 2D case. On the other hand, the scatterers
in the type of crystals used by us are not in contact which could
enable to introduce a source inside the crystal \cite{inside2D} without
any additional perturbation. This should enable the characterization
of the directional propagation inside the 3D crystal, which is not
possible in the close-packed spheres structures, where a defect should
be introduce in the structures to locate the emitter inside. 

This work has been supported by the Spanish MICINN and the European
Union FEDER through Project FIS2008-06024-C03 and -C02 and by Universitat
Politècnica de València through Project 20080025. I. P.-A. acknowledges
financial support from Generalitat Valenciana through the program
``Ajudes per a estades de doctors en centres d'excel·lència de la
Comunitat Valenciana''. 

\pagebreak{}

\pagebreak{}

\part*{Figure Captions}

\noindent Fig.1- (a) Unit cell scheme and (b) photograph of the crystal
used in the experimental setup.\\

\noindent Fig. 2- Transverse profile of the ultrasonic beam measured
at (a) 3 mm from the trasducer, (b) at 115 mm from the trasducer in
free propagation and (c) at the crystal output located at 115 mm from
the trasducer, after propagating through the crystal.\\

\noindent Fig. 3. Beam width versus frequency, dots represent the
experimental points with the corresponding dispersion bars, and the
analytical fit (see text below) is depicted by the continous line. 
\end{document}